\documentclass[epjc3]{svjour3}
\smartqed  % flush right qed marks, e.g. at end of proof
\usepackage{amsmath,amsfonts,graphics,epsfig,amssymb,color}
\newcommand{\fk}{F_{K}(Q^{2})}
\newcommand{\red}{\protect\color{red}}
\newcommand{\blue}{\protect\color{blue}}
\newcommand{\black}{\protect\color{black}}
\renewcommand\blue\red
\renewcommand\red\black
\journalname{Eur. Phys. J. C}

\begin{document}
\title{%
The $K$-meson form factor and charge radius: linking low-energy data to
future Jefferson Laboratory measurements}

\author{A.F.~Krutov\thanksref{e1,addr1}
        \and
        S.V.~Troitsky\thanksref{e2,addr2}
        \and
        V.E.~Troitsky\thanksref{e3,addr3}%etc.
}
\thankstext{e1}{e-mail: krutov@ssau.ru}
\thankstext{e2}{e-mail: st@ms2.inr.ac.ru}
\thankstext{e3}{e-mail: troitsky@theory.sinp.msu.ru}
\institute{Samara University, 443086 Samara, Russia \label{addr1}
           \and
           Institute for Nuclear
Research of the Russian Academy of Sciences, 60th October Anniversary
Prospect 7a, Moscow 117312, Russia \label{addr2}
           \and
           D.V.~Skobeltsyn Institute of Nuclear Physics,
M.V.~Lomonosov Moscow State University, Moscow 119991, Russia\label{addr3}
}
%\date{October 16, 2016; v2: December 23, 2016}
%\date{}
\maketitle

\begin{abstract}
Starting from a successful model of the $\pi$-meson electromagnetic form
factor, we calculate the similar form factor, $\fk$, of the charged $K$
meson for a wide range of the momentum transfer squared, $Q^{2}$. The only
remaining free parameter is to be determined from the measurements of the
$K$-meson charge radius, $r_{K}$. We fit this single parameter
to the published
data of the NA-7 experiment which measured $\fk$ at $Q^{2}\to 0$ and
determine our preferred range of $r_{K}$, which happens to be close to
recent lattice results. Still, the accuracy in the determination of
$r_{K}$ is poor. However, future measurements of the $K$-meson
electromagnetic form factor at $Q^{2}\lesssim 5.5$~GeV$^{2}$, scheduled in
Jefferson Laboratory for 2017, will test our approach and will reduce the
uncertainty in $r_{K}$ significantly.
\end{abstract}
%%%%%%%%%%%%%%%%%%%%%%%%%%%%%%%%%%%%%%%%%%%%%%%%%%%%%%%%%%%%%%%%%%%%%%%

\section{Introduction and outline}
\label{sec:intro}
Quantitative description of particle systems in the strong-coupling regime
remains one of the most challenging problems of contemporary particle
theory. The electromagnetic structure of light mesons represents an ideal
testbed for various approaches to practical calculations at strong
coupling, including low-energy effective theories of strong interactions
and their connection to Quantum Chromodynamics (QCD). Not
surprisingly, experimental understanding of various aspects of the meson
structure comprises an important part of the scientific program of the
upgraded Jefferson Laboratory (JLab)~\cite{JLABupgrade}, presently ready
for its start. In particular, the E12-09-011 experiment, scheduled in JLab
Hall~C for 2017, will attempt to measure the $K$-meson electromagnetic
form factor, $\fk$, at the momentum transfer squared up to $Q^{2}\sim
5.5$~GeV$^{2}$ \cite{JLabK}. In this work, we address some important
implications of this expected result and revisit previous scarce data on
$\fk$ in the frameworks of a successful theoretical model.

Some time ago, a model for the electromagnetic form factor of the charged
$\pi$ meson, $F_{\pi}(Q^{2})$, has been developed (see
Refs.~\cite{KrTr-PRC2002,KrTr-PRC2003,EChAYa2009} for a detailed
description), possessing a few free parameters, fixed in
1998~\cite{KrTr-EurPhysJ2001} from the experimental data available at that
time. The model predicted subsequent \blue JLab \black experimental results
%\red REFERENCES \black
on $F_{\pi}(Q^{2})$ surprisingly
well~\cite{KrTr-PRC2009} without further tuning of parameters, despite the
fact that the experimentally accessible range of $Q^{2}$ was extended by
an order of magnitude~\cite{exp-data}. Moreover, the model with the very
same parameters predicts the correct QCD asymptotics \blue of
$F_{\pi}(Q^{2})$ \black at large $Q^{2}$~\cite{KrTr-asymp,PRD}.
Subsequently, the model was also applied to the calculation of electroweak
parameters of the $\rho$ meson~\cite{rho}, for which particular
interesting relations have been obtained. The theoretical grounds for the
model include a relativistic-invariant approximation~\cite{KrTr-PRC2002}
to the instant form of the Relativistic Hamiltonian Dynamics (see e.g.\
Ref.~\cite{KeisterPolyzou}), while the model's quantitative parameters in
the light-quark sector are fixed from the \blue successful \black
$\pi$-meson study~\cite{KrTr-EurPhysJ2001}.

\blue
The two principal benefits of our model have been demonstrated in the
$\pi$-meson study. The first one is its \textit{predictivity}: provided
the experimental value of the decay constant is fixed, only one parameter
remains to fit the mean square radius. Any further dependence on the
model details, e.g.\ in particular the choice of the
phenomenological wave function, is
negligible~\cite{KrTr-EurPhysJ2001}. The second advantage is
\textit{matching with QCD predictions} in the ultraviolet limit:
when constituent-quark masses are switched off, as expected at high
energies, the model reproduces correctly not only the functional
form of the QCD asymptotics, but also the numerical coefficient, see
Refs.~\cite{KrTr-asymp,PRD} for details. To our best knowledge,
this is the only available low-energy model reproducing the QCD
limit without introducing additional parameters. Analytical
properties of the pion form factor, as a
function on the complex $Q^{2}$ plane, obtained in our model, are the same
as follow from the general quantum field theory principles \cite{Nefedov}.
Since the model predicted successfully the values of $F_{\pi}(Q^{2})$ up
to $Q^{2}\sim 2.5$~GeV$^{2}$, we expect that it can be used for $\fk$ at
least in the same domain of momentum transfers. Our model shares its
limitations with other constituent-quark models of mesons: at present,
their parameters cannot be consistently derived from QCD without
additional experimental input.

\black

In this paper, the model is applied to the $K$ meson. This brings two
additional parameters into the game, one being the mass of the strange
constituent quark, $M_{s}$, and another describing the interaction in the
light-heavy quark system. In the case of the $\pi$ meson, the two
corresponding parameters
%\blue (as well as other constituent-quark
%parameters) \black
were uniquely determined from two measured observables,
the $\pi$-meson decay constant, $f_{\pi}$, and charge radius, $r_{\pi}$,
which determines the form-factor behaviour at low $Q^{2}$. We will
demonstrate below that, for the $K$ meson, the decay constant, $f_{K}$,
fixes one combination of parameters, while the charge radius, $r_{K}$, is
known with large uncertainties, which makes it difficult to use it for
fixing the remaining parameter. We therefore keep it free and obtain a
range of the model parameters consistent with the present data. We address
old measurements of $\fk$ at $Q^{2}\to 0$ obtained in the NA-7 experiment
at CERN SPS \cite{AmendoliaK}, which represents the most precise source of
experimental input for determination of $r_{K}$. The value of $r_{K}$
obtained in Ref.~\cite{AmendoliaK} and, since then, extensively used in
numerous experimental and theoretical works, was estimated from fitting
the data points in the pole approximation. We demonstrate that the range
of $Q^{2}$ studied in Ref.~\cite{AmendoliaK} was sufficiently large for
deviations from this approximation to become important. We fit the NA-7
data points with our exact functions for $\fk$, which results in a
slightly shifted value of $r_{K}$. We note that the obtained allowed range
of $r_{K}$, bounded by the 68\% CL agreement with the data and by the
consistency of the model, is in a better agreement with the recent lattice
results \cite{lattice} than the original result of Ref.~\cite{AmendoliaK}.

Turning to higher energies, we use the constraints on $r_{K}$ to fix the
allowed range of $\fk$ functions at modestly large $Q^{2} \lesssim
6$~GeV$^{2}$. We observe that the spread of the theoretical curves exceeds
considerably the expected precision of the E12-09-011 experiment in JLab.
Therefore, within our approach, the E12-09-011 results might be used not
only to study  $\fk$ at moderate $Q^{2}$ but also to
constrain     its behavior at $Q^{2}\to 0$ and to further narrow the
experimentally allowed range of $r_{K}$. This improvement in the value
of a very soft parameter represents an unexpected application of the coming
experiment, aimed presumably at the studies at much higher momentum
transfers, $Q^{2} \sim (0.5-5.5)$~GeV$^{2}$.

The rest of the paper is organized as follows. In
Sec.~\ref{sec:formfactor}, we describe briefly the model for $\fk$, with
the emphasis on the differences between the $\pi$- and $K$-meson models
and on the two new parameters we have to introduce. Section~\ref{sec:r_K}
addresses the experimental constraints on $r_{K}$. The NA-7 data are
reanalized here in the frameworks of our model.
In Sec.~\ref{sec:JLab}, we present the
expected $\fk$ behavior at larger $Q^{2}$ and demonstrate how the
E12-09-011 JLab experiment may improve the precision of the $r_{K}$
measurement. We briefly conclude in Sec.~\ref{sec:concl} and present a
more detailed description of the model in the Appendix.

\section{The $K$-meson electromagnetic form factor}
\label{sec:formfactor}
The approach we used is based on the instant-form Dirac Relativistic
Hamiltonian Dynamics (see e.g.\ Ref.~\cite{KeisterPolyzou}) supplemented
by the relativistic-invariant modified impulse
approximation~\cite{KrTr-PRC2002}. The form factor of a system of two
quarks with different masses\footnote{The $\pi$ meson is well described
with $M_{u}=M_{d}$.} has been calculated, within this approach, in
Ref.~\cite{BalandinaK}. For completeness, all necessary formulae are
collected in the Appendix.

In general, we need to know five parameters to proceed with the real
numerical calculation. They include
the masses of the two constituent quarks, $M_{u}$ and $M_{s}$; the
parameter of the two-quark phenomenological wave function, $b$, with the
physical meaning of the confinement scale; and two anomalous magnetic
moments of quarks, $\kappa_{u}$ and $\kappa_{s}$. The values of the latters
are fixed in the same way as it was done for the $\pi$ meson, that is
through the Gerasimov sum rules, see Ref.~\cite{kappa} for details and
the Appendix for explicit expressions. The value of $\kappa_{u}$ is,
clearly, the same as it was used for the $\pi$ meson. We also take
advantage of the working model for $F_{\pi}(Q^{2})$ where
$M_{u}=0.22$~GeV was fixed. The phenomenological confinemenet scale,
$b$, may in principle be different for different systems, and, for the
moment, we keep it as a free parameter, together with $M_{s}$.

\blue
At this point, we note that we do not vary the shape of the
phenomenological wave function $u(k)$ used in the calculation, but fix it
instead from the $\pi$-meson study and leave only the scale $b$ as a free
parameter. In early theoretical studies of our
model~\cite{BalandinaK,f_K}, various wave functions have been used, which,
in general, resulted in different predictions. However, once the model was
applied to phenomenology, the dependence on the wave function choice was
found negligible, provided the value of the meson decay constant was
fixed~\cite{KrTr-EurPhysJ2001}. The theoretical systematic uncertainty
related to the choice of the wave-function shape is small compared to the
experimental uncertainties. Note that the weak dependence of the results
of calculation of electromagnetic form factors of pseudoscalar mesons on
the shape of the wave function of constituent quarks has been pointed out
also in Ref.~\cite{Schlumpf}.

The choice of the light constituent quark mass $M_{u}$ was determined from
a fit to experimental values of $F_{\pi}(Q^{2})$ at small $Q^{2}$ in
Ref.~\cite{KrTr-EurPhysJ2001} and confirmed by subsequent experimental
data. It is interesting to note that, with the same value of $M_{u}$, the
mass spectrum of light mesons had been successfully described in
Ref.~\cite{mass-spectrum}, within a different framework. We leave the
study of the meson masses in our model for future work.

Within our approach, we build up a consistent phenomenologically
successful global fit of electroweak properties of light mesons. The same
parameters have been used first for the best-studied $\pi$
meson~\cite{KrTr-EurPhysJ2001,PRD}, then for certain properties of the
$\rho$ meson~\cite{rho}; now we proceed with the $K$ meson.

\black

To fix the two free parameters, $b$ and $M_{u}$, for the $\pi$ meson, two
experimental observables were used. One was the meson decay constant,
$f_{\pi}$, and another was the meson charge radius, $r_{\pi}$, determined
from experimental measurements of  $F_{\pi}(Q^{2})$  at $Q^{2}\to 0$. For
the $K$ meson, we may equally well use the decay constant, $f_{K}=(0.1562
\pm 0.0010)$~GeV \cite{PDG}, to eliminate one of the parameters. The
expression relating $f_{K}$ to the model parameters was derived in
Ref.~\cite{f_K} and is presented in the Appendix.

It would be natural to use the experimental information on the $K$-meson
charge radius, $r_{K}$, to fix the single remaining free parameter and to
predict the behaviour of $\fk$ in the yet unexplored domain of large
$Q^{2}$. However, as we will see in the next section, this approach is
limited by the poor experimental knowledge of $r_{K}$.

\section{Experimental constraints on the $K$-meson charge radius}
\label{sec:r_K}
\begin{figure}
\centering
\includegraphics[width=0.67\columnwidth]{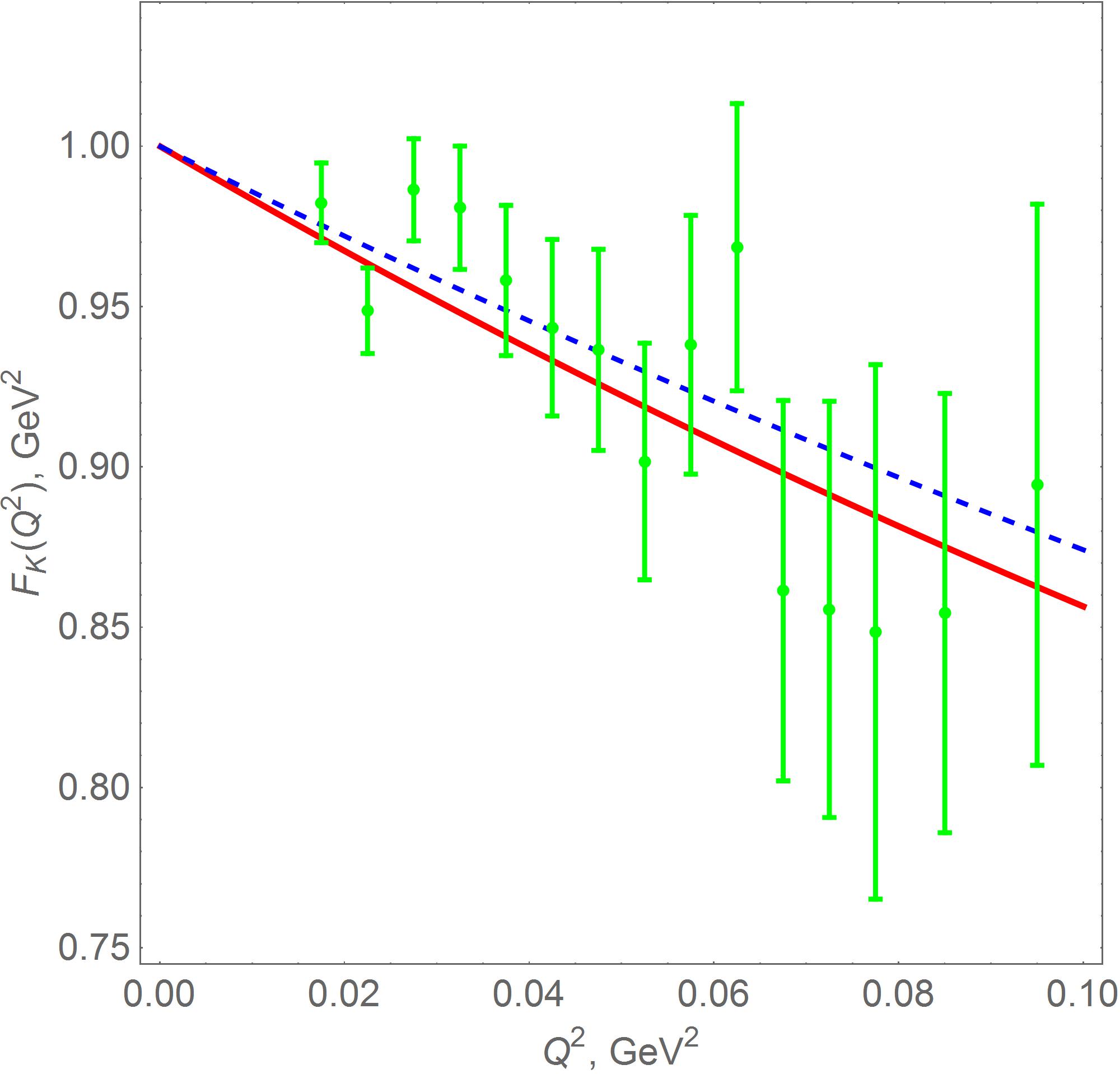}
\caption{
The NA-7 measurement of $\fk$ at low $Q^{2}$. Data points from
Ref.~\cite{AmendoliaK}. Blue dashed line: the pole fit of
Ref.~\cite{AmendoliaK}. Red full line: the best-fit curve (this work).
%For comparison, the best fit of Ref.~\cite{AmendoliaK} with the physical
%condition $F_{K}(0)=1$ relaxed is shown by the blue dotted line.
}
\label{fig:Amendolia}
\end{figure}

The form factor $\fk$ was measured by the NA-7 experiment at CERN SPS,
Ref.~\cite{AmendoliaK}. This measurement of 1986 remains the most recent
and the most precise one, and we will concentrate on its results in what
follows\blue\footnote{\blue Inclusion of an earlier measurement
of Ref.~\cite{Dally} cannot change our result because of larger
error bars and smaller number of data points, all of which lay farther away
from $Q^{2}=0$.}\black. Figure~\ref{fig:Amendolia}
presents the experimental data points. The authors of
Ref.~\cite{AmendoliaK} used these data to extract the $K$-meson charge
radius by fitting their data with the pole approximation,
$$
\fk=1/\left(1+Q^2 \langle r_K^2 \rangle/6   \right)
$$
(we note in passing that a better fit to data points was obtained in
Ref.~\cite{AmendoliaK} when the condition $F_{K}(0)=1$ was not used,
though a departure from this condition is unphysical). This resulted in
the value of $r_{K}^{2}=0.34 \pm 0.05$~fm$^{2}$ (50\% CL), widely used in
subsequent studies.

\begin{figure}
\centering
\includegraphics[width=0.67\columnwidth]{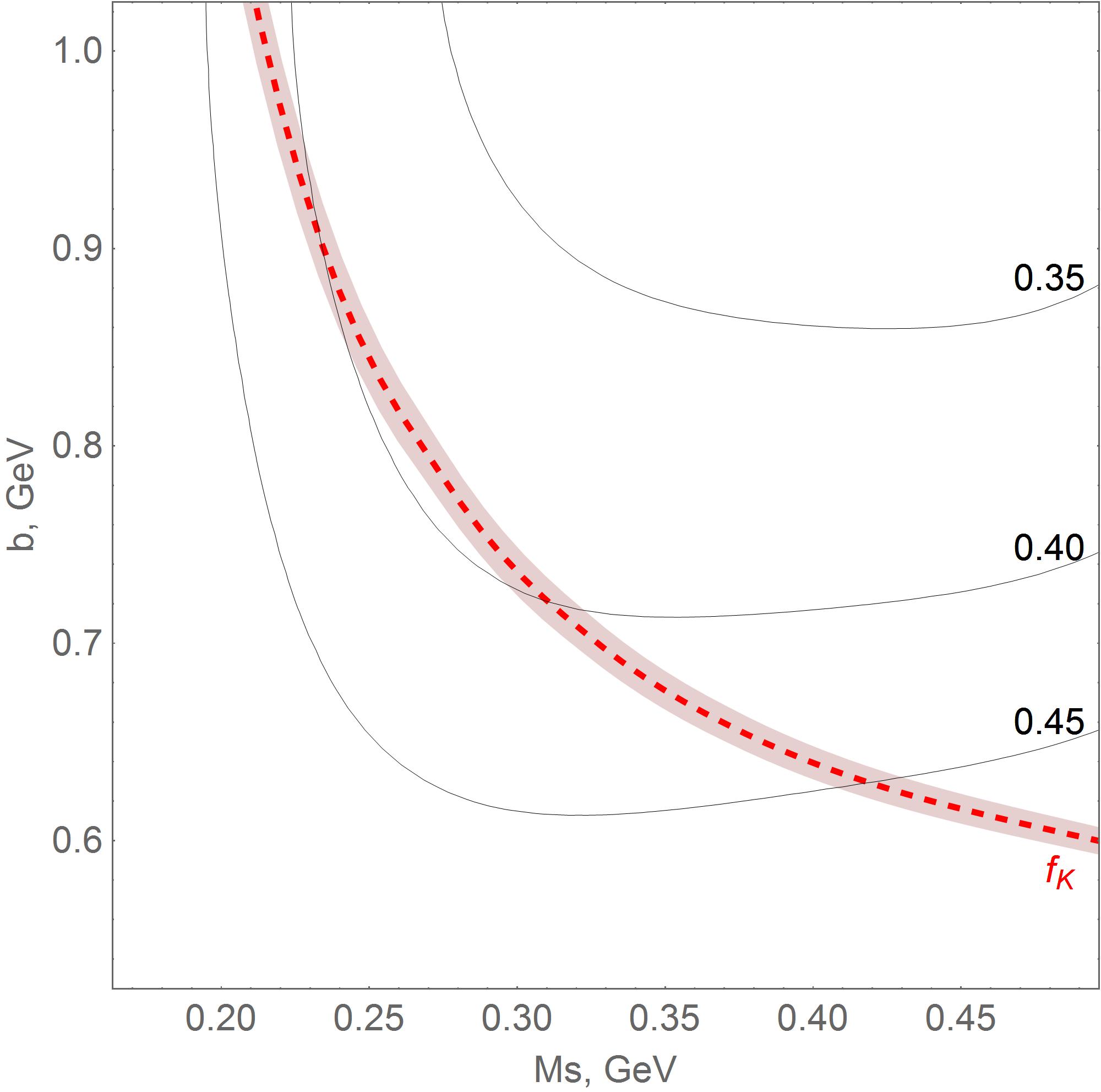}
\caption{
The two parameters for the $K$-meson form factor. The experimental value
of $f_{K}$ is reproduced on the red dashed line, with the pink shade
representing the experimental uncertainty (this condition leaves
essentially one-dimensional parameter space). Thin black lines correspond
to different values of $r_{K}^{2}$, indicated by numbers (in fm$^{2}$).
%The blue shaded area to the left is excluded by the condition
%$M_{s}>M_{u}$.
}
\label{fig:Ms-b}
\end{figure}

\begin{figure}
\centering
\includegraphics[width=0.67\columnwidth]{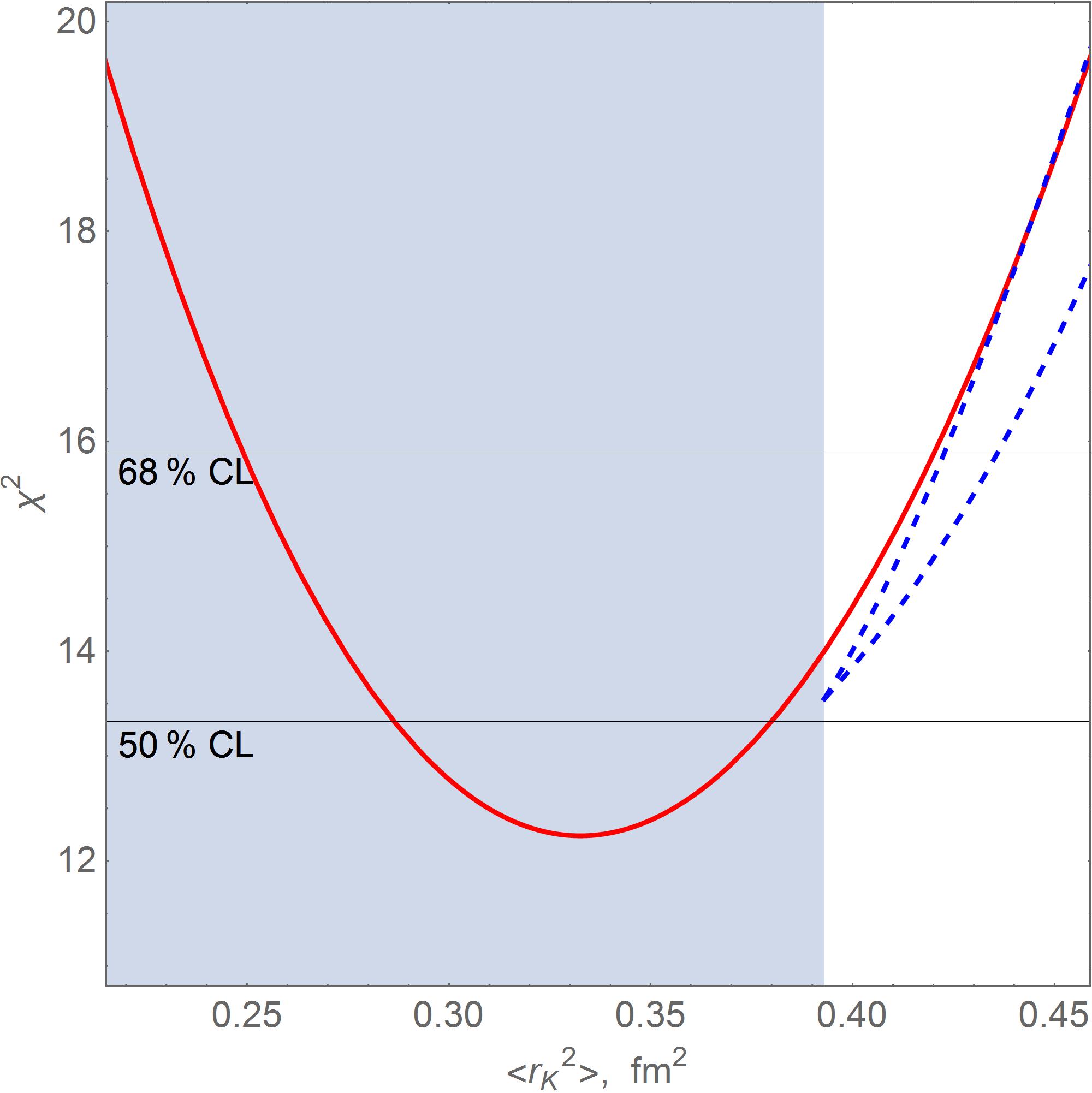}
\caption{
The $\chi^{2}(r_{K})$ function for our fits (blue dashed lines for two
branches, see text).
%; the lowest curve is cut by the condition
%$M_{s}<M_{u}$).
The shaded range, $r_{K}^{2}<0.39$~fm$^{2}$, is not
compatible with our model. For comparison, the $\chi^{2}(r_{K})$ function
for the pole fit of Ref.~\cite{AmendoliaK} is shown by the full red line.
Horisontal lines bound the 50\% CL (used in Ref.~\cite{AmendoliaK}) and
68\% CL ranges. }
\label{fig:chi2}
\end{figure}

However, one may note that the actual $\fk$ function may deviate from the
pole approximation
already at $Q^{2}\sim 0.1$~GeV$^{2}$, so that
corrections to the pole approximation are already important for the NA-7
data range. This means that, to determine the derivative of $\fk$ at
$Q^{2}=0$, and hence $r_{K}$, one should use either a shorter range of
$Q^{2}$ or a more complicated approximation. The first option fails
because of the insufficient number of data points. Fortunately, as
described above, we can calculate the function $\fk$ within our approach.
The model has one free parameter which we use to fit, by means of the
usual $\chi^{2}$ method, the NA-7 data points \blue with their
experimental error bars\black. To do that, we consider the two-dimensional
parameter space $(M_{s},b)$ of the model, see Fig.~\ref{fig:Ms-b}.
Fixing the value of $f_{K}$ implies a constraint on $(M_{s},b)$, so that a
one-parametric space remains (one can see from Fig.~\ref{fig:Ms-b} that
the precision of the experimental value of $f_{K}$ is so good that its
uncertainty may be neglected). The remaining freedom is therefore
parametrized by a line on the $(M_{s},b)$ plane; different points on the
line correspond to different values of $r_{K}$. Changing $M_{s}$ and
always keeping $b(M_{s})$ to satisfy the $f_{K}$ constraint, one may fit
the experimental data points for $\fk$. One may note, however, from
Fig.~\ref{fig:Ms-b}, that not all values of $r_{K}$ may be achieved,
provided the $f_{K}$ constraint is satisfied.
\blue Indeed, at $\langle r_K^{2}(M_{s},b) \rangle < 0.39$~fm$^{2}$, the
two curves determined by $f_{K}(M_{s},b)$ and $r_{K}(M_{s},b)$,
Fig.~\ref{fig:Ms-b}, have no intersection points. Since all other
parameters beyond $M_{s}$ and $b$ are fixed from the $\pi$-meson studies
and their values are confirmed experimentally, we have no freedom to
change this picture. Therefore, the limitation in simultaneous description
of $f_{K}$ and $r_{K}$ is a consiquence of our requirement of a
joint description of $\pi$ and $K$ mesons, and not of the
construction of our relativistic model, whose predictivity is
manifested in this way. \black The value of $M_{s}\approx 0.27$~GeV
corresponds to the lowest achievable $r_{K}$, which splits the $b(M_{s})$
line into two branches, so that a larger value of $r_{K}$ may be obtained
for two distinct values of $M_{s}$.
%The low-$M_{s}$ branch is
%additionally constrained by the condition $M_{s}>M_{u}$ which we impose.
In Fig.~\ref{fig:chi2},
we illustrate the results of the fit by presenting
the
$\chi^{2}(r_{K})$ function determined by this method.
We note that the best fit \blue ($\chi^{2}=13.5$ for 14 degrees of freedom)
\black corresponds to the lowest value of $r_{K}$, allowed in our
approach, and that
\begin{equation}
0.39~\mbox{fm}^2 \le \langle r_K^2 \rangle \le 0.42~\mbox{fm}^2~\mbox{(68\%
CL, this work)}
\label{*}
\end{equation}
Our best-fit
curve is also presented in Fig.~\ref{fig:Amendolia}.
It is interesting to note that our \blue best-fit \black value of
\blue $\langle r_{K}^{2} \rangle=0.39$~fm$^{2}$ \black
is in a better agreement with the recent lattice
calculations~\cite{lattice}, $\langle r_{K,\rm lattice}^{2} \rangle= 0.380
\pm 0.033$~fm$^{2}$, than the \blue best-fit \black value derived with the
pole approximation. However, we note that, being derived from the same
data, our value for $r_{K}$ shares similar large statistical uncertainties
with the original result\blue, and both are compatible at 68\% C.L. It is
interesting to compare our result also with the values obtained from the
data analysis in the frameworks of the chiral perturbation
theory~\cite{Bijnens}, which vary between
$\langle r_{K, \rm ChPT, min}^{2} \rangle=0.354 \pm 0.071$~fm$^{2}$
and $\langle r_{K, \rm ChPT, max}^{2} \rangle=0.431 \pm 0.071$~fm$^{2}$:
our
68\% C.L.\ interval for $\langle r_K^2 \rangle$ is contained in that of
Ref.~\cite{Bijnens} for all their assumptions.
\black

\section{Large $Q^{2}$ and the future JLab experiment}
\label{sec:JLab}
\begin{figure}
\centering
\includegraphics[width=0.67\columnwidth]{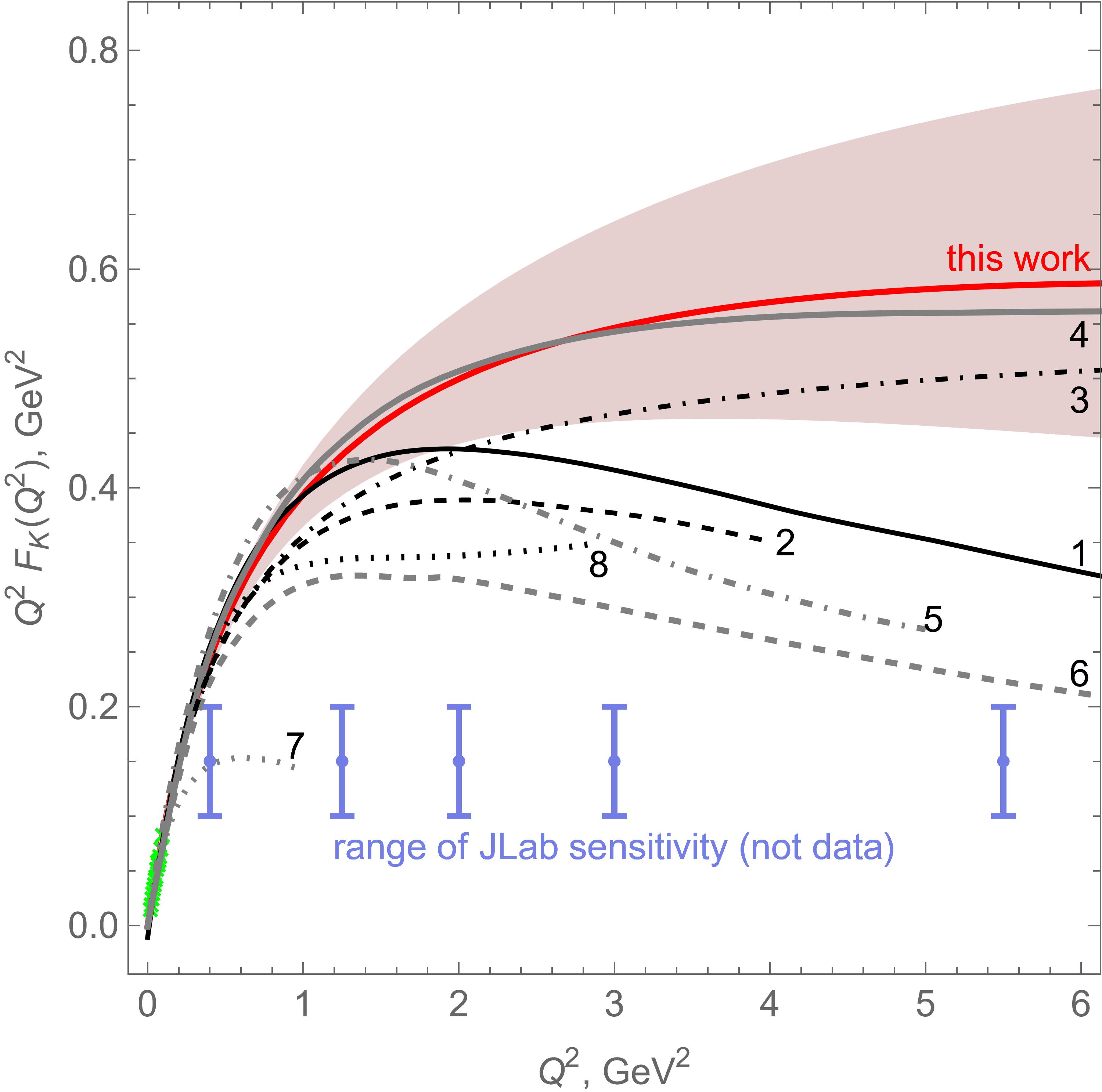}
\caption{
Predictions of our model for $\fk$ are parametrized by a single free
parameter which we fit to the low-energy data of Ref.~\cite{AmendoliaK}
(green crosses at $Q^{2}\to 0$). The red line represents the best fit and
the pink shaded area represents the 68\% CL range.
% (with an additional
%constraint $M_{s} \ge M_{u}$).
Also shown are predictions
obtained in other works. 1 (full
black): Ref.~\cite{1}; 2 (dashed black): Ref.~\cite{2}; 3 (dot-dashed
black): Ref.~\cite{4}; 4 (full gray): Ref.~\cite{5}; 5 (dot-dashed gray):
Ref.~\cite{6}; 6 (dashed gray): Ref.~\cite{7}; 7 (dotted gray):
Ref.~\cite{8}; 8 (dotted black): Ref.~\cite{Galkin}. Points with error
bars
are not data points, they represent the $Q^{2}$ range and projected
uncertainties of the E12-09-011 JLab experiment~\cite{HornRoberts2016}.}
\label{fig:FK-comparison}
\end{figure}

The use of the low-$Q^{2}$ data constrains\blue, by Eq.~(\ref{*}), \black
the remaining free parameter of the model through the fitting procedure
described in the previous section. In principle, this allows us to
calculate the $\fk$ function for a large range of $Q^{2}$ and to make
predictions for the future JLab measurements. This prediction is presented
in Fig.~\ref{fig:FK-comparison},
where the $Q^{2}$ range probed by the E12-09-011 JLab experiment and
estimated uncertainties of the measurement \cite{HornRoberts2016} are
shown. For comparison,
we present also predictions obtained within other approaches.

Considering Fig.~\ref{fig:FK-comparison}, one
immediately notes that the uncertainty in our predictions, determined by
the uncertainty in the $r_{K}$ measurements, exceeds the expected
precision of the experiment and is of order the typical difference between
model predictions. The E12-09-011 experiment would be able to
distinguish between our model and several other ones.
\blue
Since these approaches differ in their original assumptions as well as in
approximations being used, future JLab experiments will be able, in
principle, to contribute to the choice between various models of the
description of nonperturbative dynamics of strong interactions at large
and intermediate distances. In this context, it is instructive to compare
our results with those of Ref.~\cite{7}. Their approach differs from ours
by the choice of the form of relativistic dynamics: they use the
light-front dynamics. Another difference is in the approximations: while
we use the so-called modified impulse approximation, see e.g.\
Ref.~\cite{EChAYa2009}, the conventional impulse approximation was used in
Ref.~\cite{7}. Our modified impulse approximation, unlike the conventional
one, does not violate covariance conditions, nor the current conservation.
Another difference with Ref.~\cite{7}, which may be important at large
$Q^{2}$, is in the $Q^{2}$-dependence of quark form factors: we use a
renormalization-group inspired logarithmic function, while in
Ref.~\cite{7}, a dipole form is used. Note that both the value of $M_{u}$
and the expression for the quark radius coincide in the two works.

\black

\begin{figure}
\centering
\includegraphics[width=0.67\columnwidth]{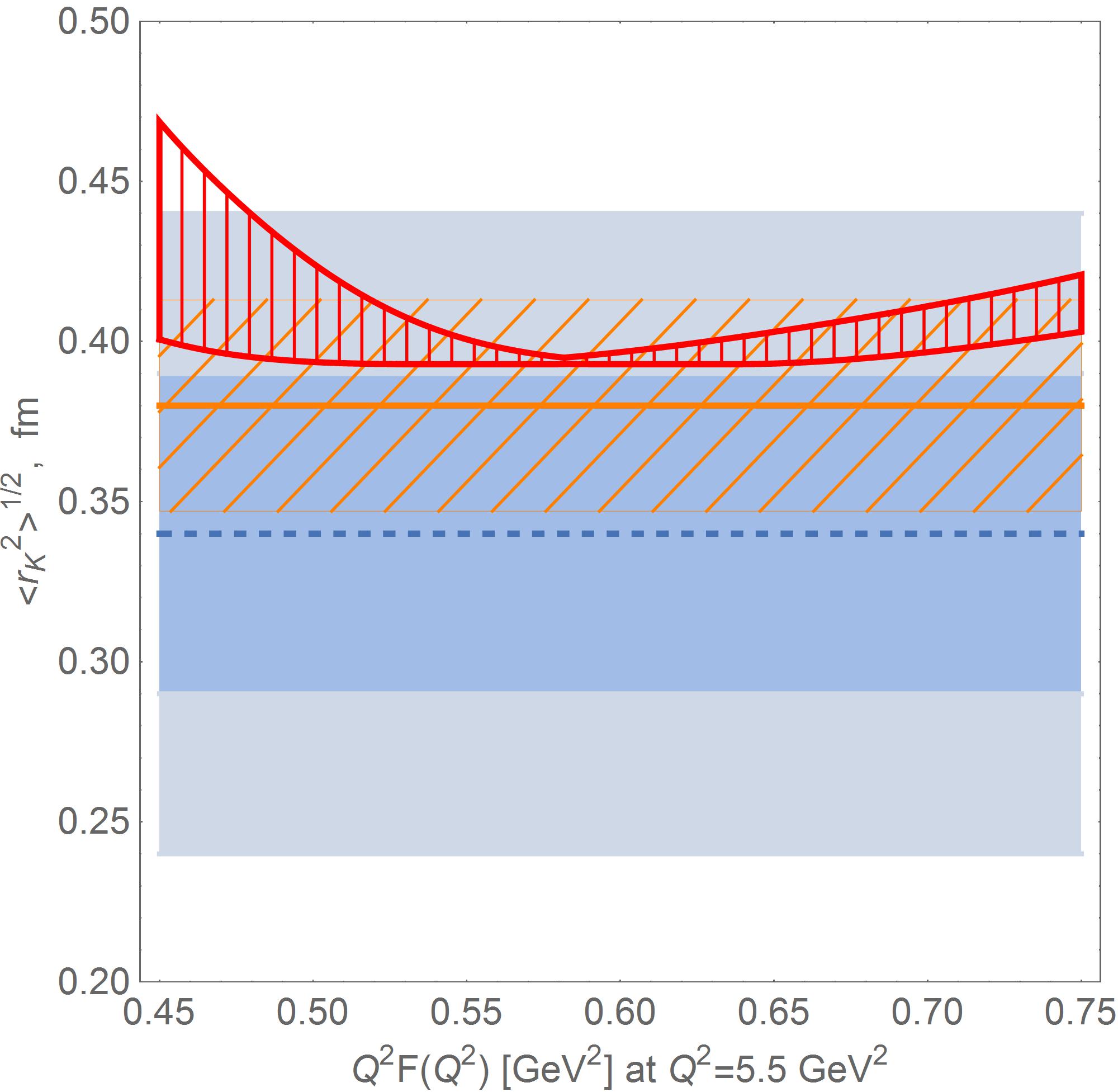}
\caption{
The effect of future JLab measurements on the precision of the $r_{K}$
value. Suppose that JLab finds the value of $F_{K}(Q^{2}=5.5~{\rm
GeV}^{2})$ shown in the horizontal axis, with the uncertainty given in
Ref.~\cite{HornRoberts2016} (conservative value). Then, within our
approach, one would be able to constrain $r_{K}$ to lay between the
lower and upper thick red curves for this value of $F_{K}$
(the strip with red vertical hatching). For comparison, $r_{K}$ of
Ref.~\cite{AmendoliaK} is shown by the dashed line (best fit), dark blue
shade (68\% CL) and light blue shade (95\% CL). The horizontal strip with
orange diagonal hatching represents the lattice result of
Ref.~\cite{lattice}.}
\label{fig:rK-from-JLab}
\end{figure}

In addition, this consideration opens a surprising new possibility to
constrain the low-$Q^{2}$ behavior of $\fk$ and to reduce the experimental
uncertainty in $r_{K}$. Indeed, the expected error bars of the experiment
are smaller than the spread of $\fk$ curves predicted in our model. This
spread is determined by \blue the 68\% C.L.\ allowed variations of the
\black single \blue not firmly fixed \black parameter of the model\blue,
$r_{K}$, \black which determines the behavior of $\fk$ at $Q^{2}\to 0$.
Hence, limiting the spread at large $Q^{2}$ would narrow the allowed range
of $r_{K}$. Figure~\ref{fig:rK-from-JLab}
illustrates how the measurement of $F_{K}(5.5~\mbox{GeV}^{2})$ in JLab
would constrain $r_{K}$ in case the uncertainty of the measurement agrees
with the estimate of Ref.~\cite{HornRoberts2016}. We note that the reduced
uncertainty in $r_{K}$ transforms, within our approach, into a more
precise knowledge of the model parameter, for instance, $M_{s}$, which may
be interesting from a theoretical point of view
(cf.\ Fig.~\ref{fig:Ms-b}).

\section{Conclusions}
\label{sec:concl}
In this work, we discussed present and future experimental data on the
$K$-meson electromagnetic form factor, $\fk$, in the context of the model
for the electorweak structure of light mesons based on the
relativistic-invariant modified impulse approximation in the frameworks of
the instant-form Relativistic Hamiltonian Dynamics. All but two parameters
of the model are fixed from its successful application to the $\pi$-meson
form factor in previous studies, where, experimentally, the model
predictions had been confirmed precisely by a number of subsequent
measurements, and theoretically, the correct QCD asymptotics was
reproduced. Of the two parameters specific for the $K$-meson case, one
combination is fixed from the $K$-meson decay constant, $f_{K}$, while the
remaining one is related to the $K$-meson charge radius, $r_{K}$. The
latter is known experimentally with large uncertainties. We revisited the
NA-7 data on $\fk$ at $Q^{2}\lesssim 0.1$~GeV$^{2}$ and found
the $r_{K}$ range allowed by the data within our model.
This range agrees well with the
recent lattice results. Still, the error bars in $r_{K}$ remain large,
which makes the predictions of the model uncertain in the large-$Q^{2}$
range to be probed by the coming E12-09-011 experiment in JLab. However,
this suggested an unexpected application of the coming JLab measurements
of $\fk$ which, despite being performed at $Q^{2}\sim(0.5-5)$~GeV$^{2}$,
would improve the accuracy of the $r_{K}$ measurements.

\appendix
\section{Formulae for the form-factor calculation}
\label{sec:app}
\subsection{The form factor of a system with two quarks of different
masses}
This form factor was calculated, in the present approach, in
Ref.~\cite{BalandinaK}.

The free form factor is given by
$$
g_0(s,Q^2,s')=\frac{\sqrt{ss'}}{\sqrt{[s^2-2s(M_{s}^2+M_u^2)+\eta^2]
[s'^2-2s'(M_{s}^2+M_u^2)+\eta^2]}}
$$
$$
\times
\frac{Q^2(s+s'+Q^2)}{2[\lambda(s,-Q^2,s')]^{3/2}}
\left (B^u(s,Q^2,s') + B^{\bar s}(s,Q^2,s')\right) ,
$$
where
$M_{q}$ is the mass of the constituent quark $q$,
$\eta = M^2_u-M^2_{s}$,
$$B^{\bar s}(s,Q^2,s') =
\left[ f_1^{(\bar s)}(s+s'+Q^2-2\eta)\cos(\omega_1
+ \omega_2)
\right.
$$
$$
\left.
-f_2^{(\bar s)}\frac{M_{s}}{2}\xi (s,Q^2,s')
\sin(\omega_1 +\omega_2 )\right ]\,\theta (s,Q^2,s') ,
$$
$$
\xi (s,Q^2,s') = \sqrt{-\lambda (s,-Q^2,s')M_{s}^2+ss'Q^2-\eta Q^2(s+s'+
Q^2)+Q^2\eta^2},
$$
$$\lambda(a,b,c)=a^2+b^2+c^2-2(ab+ac+bc),
$$
$$f_1^{(\bar s)}=\frac{2M_{s}\,G_E^{(\bar s)}(Q^2)}{\sqrt{4M_{s}^2+Q^2}};
$$
$$
f_2^{(\bar s)}=-\frac{4\,G_M^{(\bar s)}(Q^2)}{M_{s}\sqrt{4M_{s}^2+Q^2}},
$$
the Wigner rotation parameters are
$$
\omega_1=
\arctan \frac{\xi (\,s\,,Q^2\,,s')}{M_u[(\sqrt{s}+\sqrt{s'})^2+Q^2]
+(\sqrt{s}+\sqrt{s'})(\sqrt{ss'}+\eta)},
$$
$$
\omega_2=\arctan [(\sqrt{s}+\sqrt{s'}+2M_{s})\,
\xi (\,s\,,Q^2\,,s')
$$
$$
\times
\{M_{s}(s+s'+Q^2)(\sqrt{s}+\sqrt{s'}+2M_{s})+\sqrt{ss'}
(4M^2_{s}+Q^2)
-\eta [2M_{s}(\sqrt{s}+\sqrt{s'})-Q^2]\}^{-1}],
$$
$$
\theta (s,Q^2,s') = \vartheta (s'- s_1) - \vartheta (s'- s_2),
$$
$\vartheta$ is the conventional step function,
$$
s_{1,2}=M_{s}^2+M_u^2+\frac{1}{2M_{s}^2}(2M_{s}^2+Q^2)(s-M_{s}^2-M_u^2)
$$
$$
\mp\frac{1}{2M_{s}^2}\sqrt{Q^2(4M_{s}^2+Q^2)[s^2-2s(M_{s}^2+M_u^2)+\eta^2]}.
$$
The function $B^u(s,Q^2,s')$ is obtained from $B^{\bar s}(s,Q^2,s')$ by
the substitution $M_{s} \leftrightarrow M_u$ everywhere.

The quark form factors are
$$
G_E^{q}(Q^2)
=
e_q f_q(Q^2),
~~~
G_M^{q}(Q^2)
=
(e_q + \kappa_q) f_q(Q^2),
$$
where $e_{q}$ are quark charges, $f_{q}(Q^{2})=1/(1+\log(r_{q}^{2}
Q^{2}/6))$ and $r_{q}^{2}=0.3/M_{q}^{2}$.

The anomalous magnetic moments $\kappa_{q}$ of quarks $q$ are calculated
following Ref.~\cite{kappa}. The values of $\kappa_{u}$ and
$\kappa_{\bar d}$ \blue should satisfy~\cite{kappa} \black
\begin{equation}
\frac{e_u+\kappa_u}{e_{\bar d}+\kappa_{\bar d}}=-1.77\blue.\black
\label{**}
\end{equation}
\blue
The $\pi$-meson form factor depends~\cite{KrTr-EurPhysJ2001} on the sum
$\kappa_{u}+\kappa_{\bar d}$, which has been fixed in
Ref.~\cite{KrTr-EurPhysJ2001} from the condition that the
constituent-quark parameters providing a good description of the data do
not depend on the choice of the shape of the phenomenological wave
function: in this way, the value
$\kappa_{u}+\kappa_{\bar d}=0.0268$
has been found for $M_{u}=0.22$~GeV, and we use this value in the present
study. Together with Eq.~(\ref{**}), this condition determines
$\kappa_{u}$ and $\kappa_{\bar d}$ unambiguously. \black
The value of $\kappa_{s}$ is
determined \blue \cite{kappa} \black from
$$
\left(\frac{-\frac{e_u+\kappa_u}{e_{\bar d} + \kappa_{\bar d}}+
\frac{e_{\bar s}+\kappa_{\bar s}}{e_{\bar d} + \kappa_{\bar d}}}%
{1+\frac{e_{\bar s}+\kappa_{\bar s}}{e_{\bar d} + \kappa_{\bar d}}}
\right)^2=
0.42.
$$
\blue This relation, for fixed $\kappa_{u}$ and $\kappa_{\bar d}$,
determines the value of $\kappa_{\bar s}$ up to the choice of the sign at
the square root from the right-hand side. We choose the negative sign
because in the opposite case, the solution gives an unphysically
large value of $\kappa_{\bar s}$. We arrive at the values \black of
$\kappa_{u}\approx-0.01055$ and $\kappa_{\bar s}\approx-0.08099$ \blue
which are used in this study.\black

The form factor $\fk$ is given by the double integral,
$$
F_K (Q^2)=\int \! d\sqrt s\ d\sqrt{s'}\,
\varphi (k)
\,
g_0(s,Q^2,s')
\,
\varphi (k'),
$$
where
$$
\varphi(k)=\sqrt{\sqrt{s}(1 - \eta^2/s^2)}\,u(k)\,k,
$$
$$k = \sqrt{{(s^2-2s(M^2_{s}+M^2_u)+\eta^2)}/{4s}},$$
is the phenomenological wave function of the two-quark
system. Following previous studies of the $\pi$ meson, we choose
the power-law function,
$$
u(k)=N\,{(k^2/b^2 + 1)^{-3}},
$$
where the normalization $N=16 \sqrt{2/(7 \pi b^3)}$ is determined by the
condition
$$
\int_0^\infty \! dk\, k^2 u(k)^2.
$$

\subsection{The $K$-meson decay constant}
The expression for the decay constant $f_{K}$ was calculated, in the
present approach, in Ref.~\cite{f_K}. It reads:
$$
f_K=
\int_{M_s+M_u}^\infty \! d\sqrt{s}  \,
G_0(s) \varphi(s),
$$
where
$$
G_0(s)=\frac{\sqrt{3}}{2\sqrt{2}\pi\sqrt{s}}
\sqrt{(p_{s0}+M_s)(p_{u0}+M_u)}
\left(1-\frac{k^2}{(p_{s0}+M_s)(p_{u0}+M_u)}   \right) ,
$$
$$
p_{q0}=\sqrt{M_q^2+k^2}.
$$

\begin{acknowledgements}
We thank A.~Gasparian and G.~Huber for interesting information on the
coming JLab $K$-meson experiments. The work of AK was supported in part by
the Ministry of Education and Science of the Russian Federation (grant
No. 1394, state task).
The work of ST on nonperturbative description of strongly coupled QCD
bound states is supported by the Russian Science Foundation, grant
14-22-00161.
\end{acknowledgements}

\end{document}